# Temporal Signal Processing with Nonlocal Optical Metasurfaces

Michele Cotrufo[1,2,†*], Sedigheh Esfahani[1,3,†], Dmitriy Korobkin[1,†] and Andrea Alù[1,3*]

[1]Photonics Initiative, Advanced Science Research Center, City University of New York, New York, NY 10031, USA

[2]The Institute of Optics, University of Rochester, Rochester, New York 14627, USA

[6]Physics Program, Graduate Center of the City University of New York, New York, NY 10016, USA

[†]These authors contributed equally

* *mcotrufo@optics.rochester.edu, aalu@gc.cuny.edu*

*Nonlocal metasurfaces have recently enabled an ultra-compact, low-power and high-speed platform to perform analog image processing. While several computational tasks have been demonstrated based on this platform, most of the previous studies have focused only on spatial operations, such as spatial differentiation and edge detection. Here, we demonstrate that metasurfaces with temporal nonlocalities – that is, with a tailored dispersive response – can be used to implement time-domain signal processing in deeply subwavelength footprints. In particular, we show a passive metasurface performing first-order differentiation of an input signal with high-fidelity and high-efficiency. We also demonstrate that this approach is prone to scalability and cascaded computation. Our work paves the way to a new generation of ultra-compact, passive devices for all-optical computation, with applications in neural networks and neuromorphic computing.*

## Introduction

Optical computing has recently gained renewed interest[1] due to the continued improvements in nanofabrication and to the exponentially growing demand for data processing, especially in the context of neural networks[2] and neuromorphic computing[3–5]. All-optical computation offers the advantage of processing data at the speed of light while avoiding analog-to-digital conversion[6], leading to large reductions in energy consumption and latency times. While building general-purpose optical processors is still out of reach, there is a strong interest in implementing special-purpose analog optical devices[1] performing specific operations on the incoming signals, working in parallel with electronic hardware to accelerate specific tasks or lowering the burden on digital processing. For example, neural networks and neuromorphic computing necessitate massively parallel signal processing, including detection of temporal variations of signals, and they may largely benefit from the possibility of offloading some of these tasks into high-speed analog optical devices with zero energy consumption and ultrafast speeds.

Signal and image processing in the analog domain based on Fourier optics concepts has been proposed and extensively studied in the last decades[7]. The underlying idea is to decompose a signal into its spatial and/or temporal frequency components via optical elements, such as lenses and diffraction gratings, and then selectively filter these components with the aid of spatially varying masks. Notable examples of these approaches are 4f systems for spatial image processing[8] and pulse shapers for the manipulation of temporal signals [9,10]. While versatile and easy to implement, these approaches ultimately suffer from the need of using bulky optical elements, which strongly increase the overall system footprint and prevent compactification, while also being prone to alignment issues. In recent years, it has been suggested that Fourier-based image processing can be implemented in sub-wavelength planarized patterned devices, known as metasurfaces[11,12]. In particular, nonlocal metasurfaces can act directly in the Fourier space of an input signal, thus implementing on-demand filtering of the Fourier spatial components without the need for bulky optical elements. Several theoretical[11–17] and experimental[18–25] studies have indeed demonstrated that mathematical operations, such as spatial differentiation, can be imparted on the incoming wavefront using metasurfaces. The key idea of these approaches is to transform the desired spatial operation in its Fourier counterpart, and then encode it into the angle-dependent transfer function of a metasurface.

While metasurface-based Fourier processing has been extensively investigated for spatial operations and image processing, comparatively little effort has been devoted to *temporal* operations using nonlocal metasurfaces. Similar to their spatial counterpart, linear temporal operations can be implemented by transforming the desired operation into a multiplicative polynomial factor in the temporal Fourier domain, and then encoding such a factor into the frequency-dependent transfer function of the metasurface through dispersion engineering. In other words, by tailoring its *frequency* nonlocality, a metasurface can act as a compact spectral filter performing time-domain computations *without* the need for pulse shapers and diffraction gratings, within a deeply subwavelength overall footprint. Recent theoretical works have investigated temporal signal processing with graphene[26,27] or dielectric[28] structures, including a recent experimental demonstration at radio frequencies[29]. Theoretical studies have also specifically investigated metasurfaces performing linear combinations of spatial and temporal differentiation[30–34], especially geared towards the generation of spatio-temporal optical vortices[35]. Similar concepts have been demonstrated in integrated photonics[36–39] using ring resonators and Mach-Zehnder interferometers as dispersive elements, and in radio-frequency circuits[40]. Yet, an experimental demonstration of temporal differentiation and processing of time-dependent optical signals in free-space using nonlocal metasurfaces is lacking. We note that recent works have demonstrated spatio-temporal light shaping using spatially-varying metasurfaces as masks inside traditional pulse shapers[41]. These approaches are attractive because of the increased versatility of metasurfaces in engineering the complex transfer amplitude. However, the underlying need for additional diffraction gratings and lenses sets hard bounds on the minimal footprints of the overall device, and may introduce relevant alignment issues.

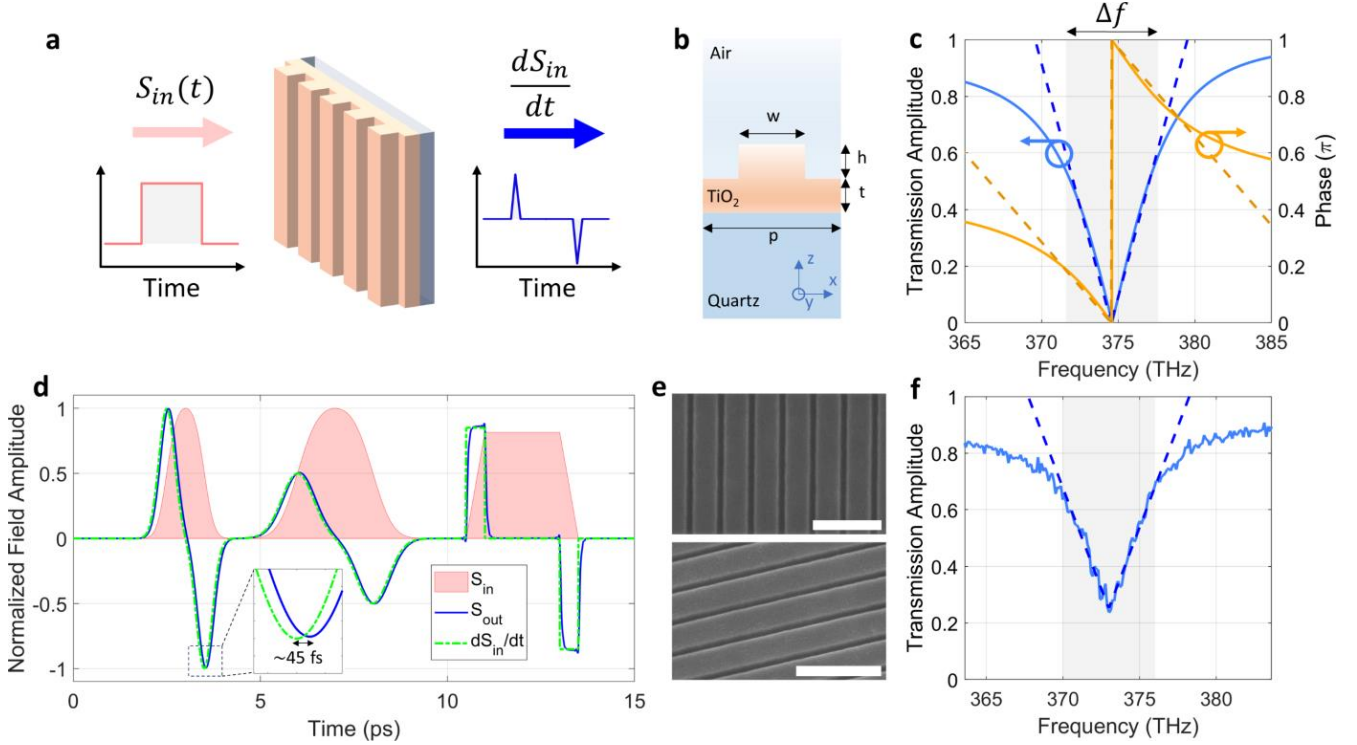

**Figure 1. General working principle and metasurface design.** (**a**) A signal $S_{in}(t)$ (e.g., a square pulse) is encoded in the envelope of an electromagnetic wave impinging on a metasurface. The envelope of the transmitted wave is the first-order derivative of the input pulse $dS_{in}(t)/dt$. (**b**) Geometry of the unit cell of the nonlocal metasurface. See text for design parameters. (**c**) Calculated amplitude (solid blue line) and phase (solid yellow line) of the transfer function of the metasurface. The dashed lines are fits of the corresponding data. Here and in all other panels the electric field is parallel to metasurface wires (*y*-direction). (**d**) Calculated response of the metasurface in panels (b-c) when a signal with amplitude $S_{in}(t)$ (red shaded area) is impinging. The blue line shows the calculated output field amplitude $S_{out}(t)$, while the green curve shows the analytically calculated value of $dS_{in}(t)/dt$. The inset shows that $S_{out}(t)$ and $dS_{in}(t)/dt$ are delayed by about 45 fs. (**e**) SEM images of a fabricated device. Scalebars = 1 µm. (**f**) Measured transmission amplitude of the fabricated device (solid line), together with a linear fit (dashed line).

Here, we demonstrate a nonlocal metasurface performing analog temporal differentiation of free-space signal $S_{in}(t)$ carried by optical waves with wavelength $\lambda_0 \approx 800$ nm (Fig. 1a). In stark contrast from approaches based on pulse shapers[9,10,41], here the computation is fully performed within a deeply subwavelength metasurface with thickness $H \approx \lambda_0/5$, without the need for additional optical elements (Fig. 1a). We achieved this functionality by tailoring the metasurface nonlocality in time, so that its transfer function follows the trend $t(\omega) \propto -i(\omega - \omega_0)$ around a certain operating frequency $\omega_0$. Such transfer function results in the first-order differentiation of input signals, modulating the impinging wave in the time domain. We experimentally characterize the response of our metasurface for different types of input pulses, and quantitatively show that the mathematical operation encoded in the transmitted signal is very close to the exact first-order derivative. As a possible application, we demonstrate that our device can be used for temporal edge detection: the filtered signal features large spikes at the same instants of time when the input signal undergoes large variations. Moreover, we quantitatively characterize the metasurface efficiency — i.e., we compare the power carried by the filtered signal with the power of the input signal — showing that the efficiency of our device is less than a factor of 2 smaller than the efficiency of any ideal passive differentiator with the same bandwidth. Finally, we experimentally demonstrate the potential of this approach for scalability and cascaded operations, showcasing how multiple metasurfaces, each performing the first-order derivative, can be cascaded without requirements on their precise positioning, resulting in an overall device that performs higher-order differentiation.

## Experimental Results

Figure 1b shows the unit cell of the optimized metasurface. It consists of a resonant waveguide grating (RWG) made of Titanium Oxide ($TO_2$) placed on a quartz substrate. A grating with lattice constant $p = 475$ nm, wire width $w = 345$ nm and depth $h = 100$ nm is partially etched into a $TO_2$ slab with total thickness $H = h + t = 160$ nm. The geometry is optimized such that the transmission spectrum of the metasurface features a dip with the desired shape at a wavelength $\lambda_0 \approx 800$ nm, corresponding to a frequency $\omega_0/2\pi = f_0 \approx 375$ THz, while simultaneously supporting no diffraction orders in either the superstrate or substrate. In the following simulations and experiments, the impinging field is polarized along the grating wires (y-direction in Fig. 1b). Figure 1c shows the calculated amplitude (solid blue line) and phase (solid orange line) of the metasurface transfer function for normal incidence. As confirmed by the simulated data (solid lines) and fits (dashed lines), within a certain range of frequencies $\Delta\omega/2\pi = \Delta f \approx 6$ THz (denoted by the grey shaded area) the transfer function can be approximated by $t(\omega) = Ae^{i\omega\tau}(\omega - \omega_0)$ (dashed lines in Fig. 1c are fits), where $\tau \approx 45\,fs$. Thus, if an electromagnetic wave with carrier frequency $f_0$ and with envelope $S_{in}(t)$ with spectrum defined within $[f_0 - \Delta f/2, f_0 + \Delta f/2]$, impinges on the metasurface from one side, the envelope of the transmitted wave is expected to be proportional to the first-order derivative of $S_{in}(t)$ delayed by $\tau$, i.e. $S_{out}(t) \propto S'_{in}(t - \tau)$. This is verified numerically in Fig. 1d. Here, we considered an input signal $S_{in}(t)$ (red shaded area) containing pulses with different shapes. We calculated the output field $S_{out}(t)$ produced by the metasurface (solid blue line) assuming the realistic transfer function shown in Fig. 1c, and we compare it to the exact first-order derivative $dS_{in}/dt$ (solid green line). The field created by the metasurface precisely follows the first-order derivative of the input field, except for a delay of $\tau \approx 45$ fs (see inset of Fig. 1d), as expected.

Following this optimization, we fabricated $TiO_2$ metasurfaces with geometrical parameters close to the simulated ones (see Methods for details). Fig. 1e shows SEM micrographs of a fabricated device. The measured normal-incidence transmission spectrum (Fig. 1f) confirms that the transmission amplitude follows the expected $|t(\omega)| \propto |\omega - \omega_0|$ dependence within a $\Delta f = 6$ THz range centered around $\omega_0/2\pi = f_0 = 373$ THz. As compared to simulations, the minimum of the measured transmission amplitude is not zero, due to fabrication imperfections and sub-idealities in our setup. Nonetheless, the minimum transmission $T = |t(\omega)|^2$ is below 7%, ensuring a good suppression of the DC component of the signal envelope.

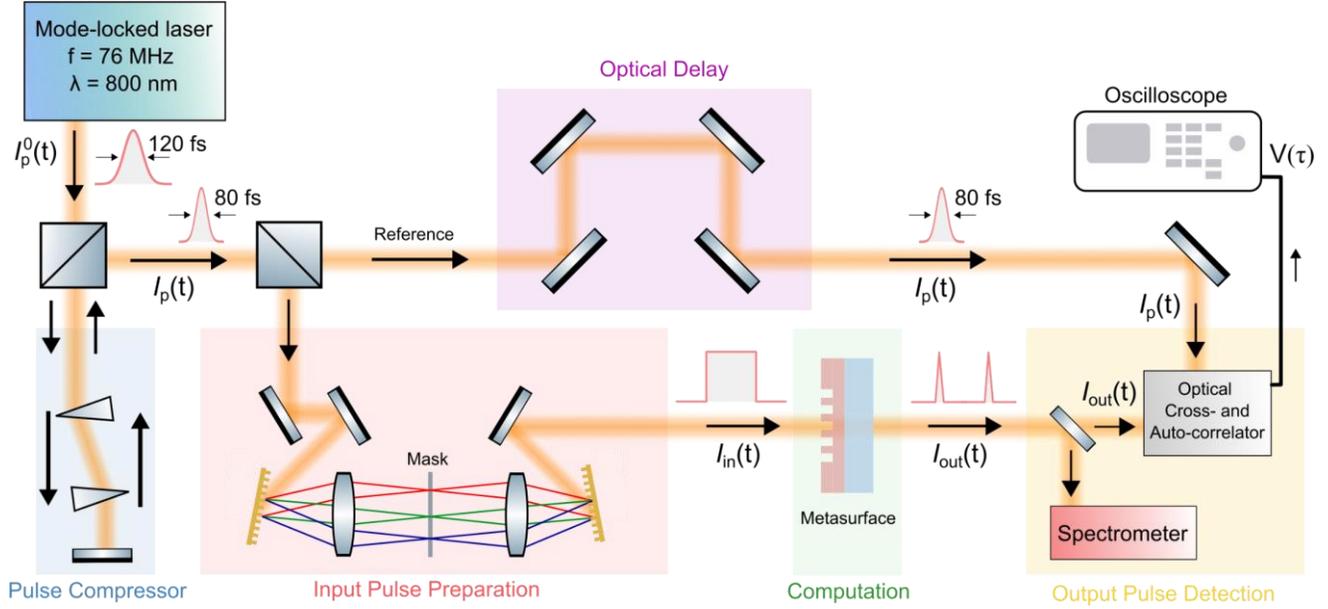

**Figure 2. Experimental setup.** The 120-fs pulses generated by a mode-locked laser are first compressed (to about 80 fs) by a prism-based pulse compressor and then split into two copies with a beam splitter. One copy, $I_p(t)$ is delayed by a controllable optical delay and then sent to the reference port of an optical cross-correlator. The other copy is redirected to a pulse shaper, which is used to generate input pulses $I_{in}(t)$ with arbitrary temporal shapes. The input pulse is then sent to the metasurface. The pulse $I_{out}(t)$ is sent either to a spectrometer or to the second port of the cross correlator. See text for additional details.

In order to experimentally verify the temporal signal processing operation of our nonlocal metasurface under different excitations, we used the custom-build setup schematized in Fig. 2. To generate input signals $S_{in}(t)$ with arbitrary temporal shapes and large spectral bandwidths, we used the output of a CW mode-locked Ti:Sapphire laser (repetition rate 76 MHz, central wavelength 800 nm) as a broadband source. The laser produces positively chirped pulses with a duration of 120 fs and a bandwidth of $\Delta\nu$ = 5.43 THz. A prism compressor[42] was used to re-compress the pulses to quasi-Fourier-transform-limited pulses with a duration $\Delta\tau \approx$ 80 fs (time-bandwidth product $\Delta\tau \cdot \Delta\nu \approx$ 0.43). Such compressed pulses [$I_p(t)$ in Fig. 2] were sent onto the input port of a beamsplitter and split into two copies. One copy of the pulse ('reference' in Fig. 2) was sent through an optical delay line, and then into the reference port of an optical cross-correlator (OCC). Another copy of the pulse was sent into a custom-built pulse shaper[9,10] (red-shaded area in Fig. 2 labelled "Input Pulse Preparation"). Here, the different spectral components of the pulse $I_p(t)$ were spatially separated by a diffraction grating and a lens, filtered by a mask, and then recombined by a second set of diffraction grating and lens. In such a way, different desired input pulse amplitudes $S_{in}(t)$ can be generated using masks with different spatially varying transmission profiles. The input pulse $S_{in}(t)$ is then directed onto the metasurface with a long-focal-length lens (f = 50 cm). The signal processed by the metasurface $S_{out}(t)$ was then collected on the other side by a second lens and sent either onto a spectrometer or onto the input port of the OCC. The voltage signal produced by the OCC is proportional to the convolution of the reference pulse intensity $I_p(t)$ and the intensity $I_{out}(t) \propto |S_{out}(t)|^2$ of the output signal, i.e., $V_{out}(\tau) \propto \int dt\, I_p(t) I_{out}(\tau - t)$. Thus, by working with narrow reference pulses $I_p(t)$, the cross-correlation trace (CCT) $V_{out}(\tau)$ allows retrieving the field amplitudes $|S_{out}(t)|$. In order to experimentally measure also the input field $S_{in}(t)$, we laterally translated the metasurface sample, so that the input field $S_{in}(t)$ impinged on an area of the sample with unpatterned 160-nm-thick $TiO_2$. Here, the spectral response of the sample is flat, and thus, apart from an intensity attenuation of less than 4%, the transmitted field is equal to $S_{in}(t)$. Therefore, the corresponding CCT $V_{in}(\tau)$ allows us to retrieve the field amplitude $|S_{in}(t)|$.

We emphasize that, while the experimental setup in Fig. 2 involves several bulk components and may appear fairly complex, the analog computation occurs only within the sub-wavelength metasurface (green-shaded area labelled "computation" in Fig. 2). All other parts of the setup are used either to prepare a certain input pulse of interest or to detect the output signal and verify the metasurface operation, but they do not play any role in the computation of the first-order derivative. In a practical setup, the signal processing metasurface can be placed anywhere in front of a detector to impart the desired mathematical operation on the impinging signals.

Leveraging the setup in Fig. 2, we have tested the response of the metasurface under different input signals. In Fig. 3a, we consider the case a rectangular pulse with duration of approximately 1 ps as the input signal. This pulse was produced using in the pulse shaper a mask imparting a transmission profile proportional to $sinc(x)$[10]. Fig. 3a shows the CCTs of the input signal (red-shaded area) and of the output signal (solid blue line), multiplied by 10. As expected from a first-order differential operation, the output signal features two high-contrast peaks at the same temporal positions where the intensity of the input signal abruptly changes from zero to a finite value or vice versa. Instead, for instants for which the intensity of the input field envelope remains approximately constant, the output signal is almost zero. The residual non-zero signal in the output field is due to the fact that the experimental transfer function (Fig. 1f) is not exactly zero at $f = f_0$, leading to a partial transmission of the DC component of the input signal envelope. The experimental results in Fig. 3a show that our metasurface performs temporal edge detection, in analogy with the spatial edge detection recently reported using nonlocal metasurfaces performing the Laplacian operation. Similar results are obtained when the input is a rectangular pulse with a longer duration of 2 ps (Fig. 3b). In this case, a larger background signal between the two peaks (solid blue line in Fig. 3b) is observed in the output signal. This feature is expected since the input signal bandwidth is narrower in this scenario, hence the output signal is affected more pronouncedly by the non-zero transmission at $f = f_0$. We also considered a rectangular pulse with asymmetric edges (Fig. 3c), obtained by intentionally misaligning the corresponding $sinc(x)$ mask in the pulse shaper. As expected, the edge with steepest slope (close to time t = - 1 ps) leads to a stronger peak in the output signal, while the other edge, close to time t = 1 ps and characterized by a gentler slope, gives rise to a much less intense signature in the output signal.

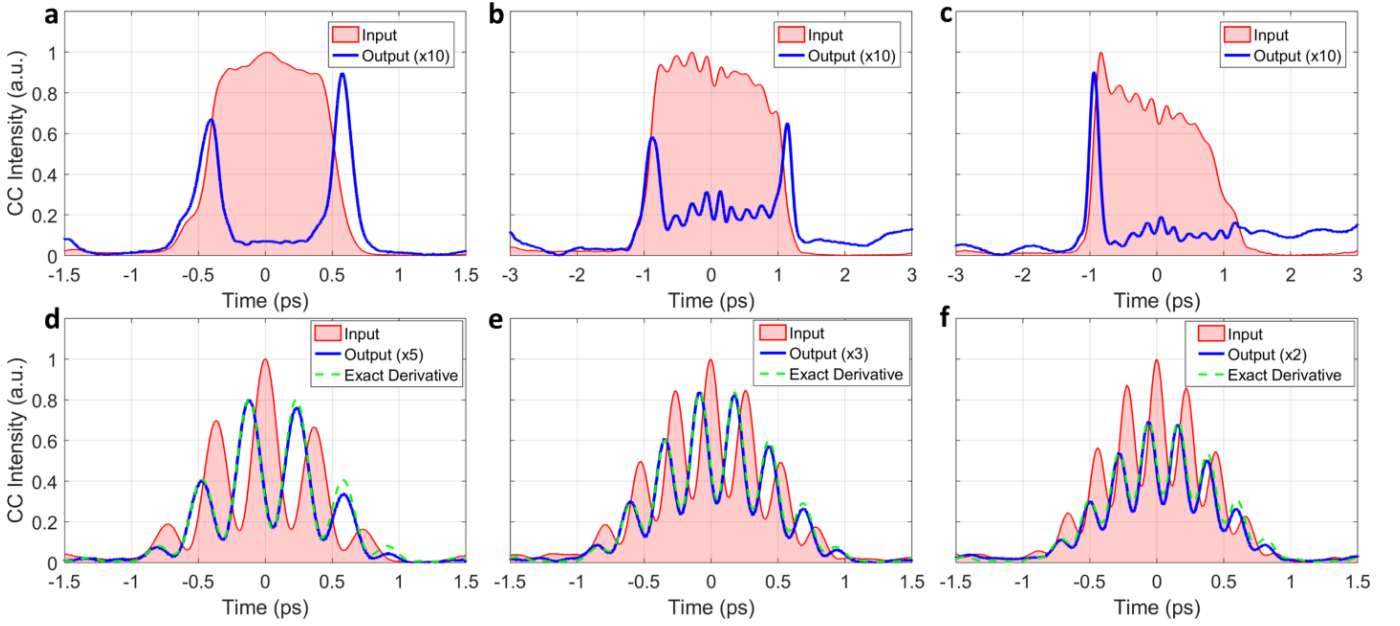

**Figure 3. Experimental measurements for different input signals.** In all panels, the red-shaded areas show the CCT of the input pulses, and the blue solid lines show the CCT of the output pulses. In panels d-g the green dashed curves show the CCT calculated for the exact first-order derivatives (see text for details). (a-c) The input pulse is a rectangular pulse with duration 1 ps (panel a) or 2 ps (panel b). In panel c, we intentionally displaced the phase mask in the pulse shaper to create an asymmetric pulse. (d-e) The input pulses are obtained by the beating of two quasi-monochromatic pulses with frequency detuning $\Delta$. The detunings are (d) $\Delta = 2.65$ THz, (e) $\Delta = 3.75\ THz, (f) \Delta =$ 4.42 THz.

Next, we verify the computational capability of the nonlocal metasurface with more complex input signals, which also allows us to verify that the performed mathematical operation is close to the exact first-order derivative. We consider input pulses obtained by the beating of two quasi-monochromatic waves with frequencies $f_0 \pm \Delta/2$. In our experiments, these pulses were obtained by placing an opaque mask with two pinholes in the pulse shaper, to select only two narrow frequencies. By changing the distance between the pinholes, different beating frequencies $\Delta$ can be generated. Figures 3(d-f) show the experimental results for three different input pulses, with beating frequencies $\Delta = 2.65$ THz (Fig. 3d), $\Delta = 3.75$ THz (Fig. 3e), $\Delta = 4.42$ THz (Fig. 3f). In all cases, the CCTs of the input signal (red-shaded areas) show the expected cosine-like shape, with the maximum intensity at t = 0 and additional peaks separated by $T \equiv 1/\Delta$. In each measurement in Figs. 3(d-f), the output CCT (solid blue line) displays an oscillatory pattern, with maxima almost aligned with the minima of the corresponding input CCT (red-shaded area). This pattern matches the behavior expected from the squared modulus of the derivative of a cosine-like function. The measurements in Figs. 3(d-f) also confirm the presence of a systematic delay: in each measurement, the central minimum of the output CCT (solid blue lines) systematically lags behind the maximum of the corresponding input CCT. By comparing the positions of these features in all traces in Figs. 3(d-f), we estimate a delay of about $\tau_{\exp} = 47$ fs, in very good agreement with the delay $\tau = 45$ fs extracted from simulations in Fig. 1c. Finally, we compare the experimental data with the CCT theoretically expected if the output field were the *exact* first order derivative of the input field. To do so, we use a procedure described in detail in the Supplementary Information and summarized here: we first built an ansatz for the input field $S_{in}(t)$, given by the sum of two Gaussian pulses with carrier frequencies $f_0 \pm \Delta/2$, and we constructed the corresponding CCT, $V_{in}^{fit}(\tau)$. This function was then used to fit the measured CCTs of the input fields, $V_{in}(\tau)$ (red-shaded area in Figs. 3(d-f)). This allowed us to retrieve several parameters in the ansatz for $S_{in}(t)$. We then analytically calculated the delayed first-order derivative $dS_{in}(t - \tau_{\exp})/dt$ and the corresponding CCT. The CCTs obtained with this procedure are plotted as green dashed lines in Figs. 3(d-f), showcasing excellent agreement with the measured CCTs of the signal processed by the metasurface (blue lines). This provides an indirect confirmation that the signal processing performed by the metasurface closely matches the desired first-order differentiation.

After having demonstrated the capability of our metasurface to perform first-order differentiation, we are ready to investigate the metasurface computational efficiency, i.e., how the intensity of the output signal compares to the intensity of the input signal. While there is an absolute limit on the efficiency that depends on the nature of the performed operation (a derivative in time is a high-pass filter, hence it naturally is expected to suppress a portion of the input energy), the metasurface dispersion implies that the overall efficiency also depends on the bandwidth of the input signal. In particular, while we can expect differentiation for any signal whose bandwidth is equal or smaller to the frequency range for which $t(\omega) \propto -i(\omega - \omega_0)$, we expect higher efficiencies for signals whose spectrum covers this entire range of frequencies, such that the spectrum edges, corresponding to spikes triggered by fast changes in the input signals, experience close to unitary transmission. To study this effect, we used input pulses with a Gaussian-like temporal shape and continuously varying spectral bandwidths. These pulses were generated by using a slit with variable width as a mask in the pulse shaper. Figures 4(a-f) show the experimental results for six different spectral bandwidths. In each figure, the left panel shows the normalized spectra of input (red-shaded areas) and output (blue solid lines) signals. The bandwidth of the input signals (measured as FWHM of the corresponding spectrum) ranges from 5.2 THz in Fig. 4a to 1.5 THz in Fig. 4f. The spectrum of the output signals (blue lines in left panels) indeed confirms that the metasurface strongly suppresses the input signal spectrum at frequencies close to $f_0 \approx 373$ THz, as expected. The right panel of each Figs. 4(a-f) shows the measured input and output CCTs (similar to the plots in Fig. 3). In all six cases, the input signal resembles a Gaussian pulse with additional weaker tails. The output signals indeed consist of two main peaks which, apart from the expected delay $\tau_{\text{exp}} \approx 47$ fs, are located at the temporal positions where the input signal has the largest slope. To quantify the metasurface efficiency, we consider the metric $\eta_{\text{int}} \equiv \int dt V_{out}(\tau) / \int dt V_{int}(\tau)$, defined as the ratio of the integrated input and output CCTs. In the limit of very short reference pulses $I_p(t)$, this metric is equal to the ratio between the energies carried by the input and output pulses. In Fig. 4g we plot (blue circles) the efficiency $\eta_{\text{int}}$ calculated for all measurements in Figs. 4(a-f) (and for other measurements not shown here) as a function of the spectral bandwidth of the input signal. As clearly shown from this plot, the efficiency increases with the bandwidth of the input pulse, as expected, ranging from $\eta_{\text{int}} = 1.4\%$ for a bandwidth of 1.2 THz to $\eta_{\text{int}} = 24\%$ for a bandwidth of 5.25 THz. As mentioned, the increase of efficiency with bandwidth is expected because, as the spectral bandwidth of the input pulse increases, its Fourier spectrum captures more and more the high-transmission regions of the metasurface transfer function (Fig. 1f). As the input spectral bandwidth increases further and becomes larger than the operational bandwidth of the nonlocal metasurface ($\Delta f = 6$ THz for this device, see also Fig. 1f), we expect to observe unwanted distortions from the expected differentiation operation. We stress again that the measured efficiency is not associated to a poor implementation of the metasurface, but rather to the fact that a derivative operation implemented via passive components expectedly produces an efficiency drop. Dealing with dielectric metasurfaces, material absorption is negligible and the bulk of the power loss is actually reflected by the metasurface in an effort to filter out the DC component of the input signal.

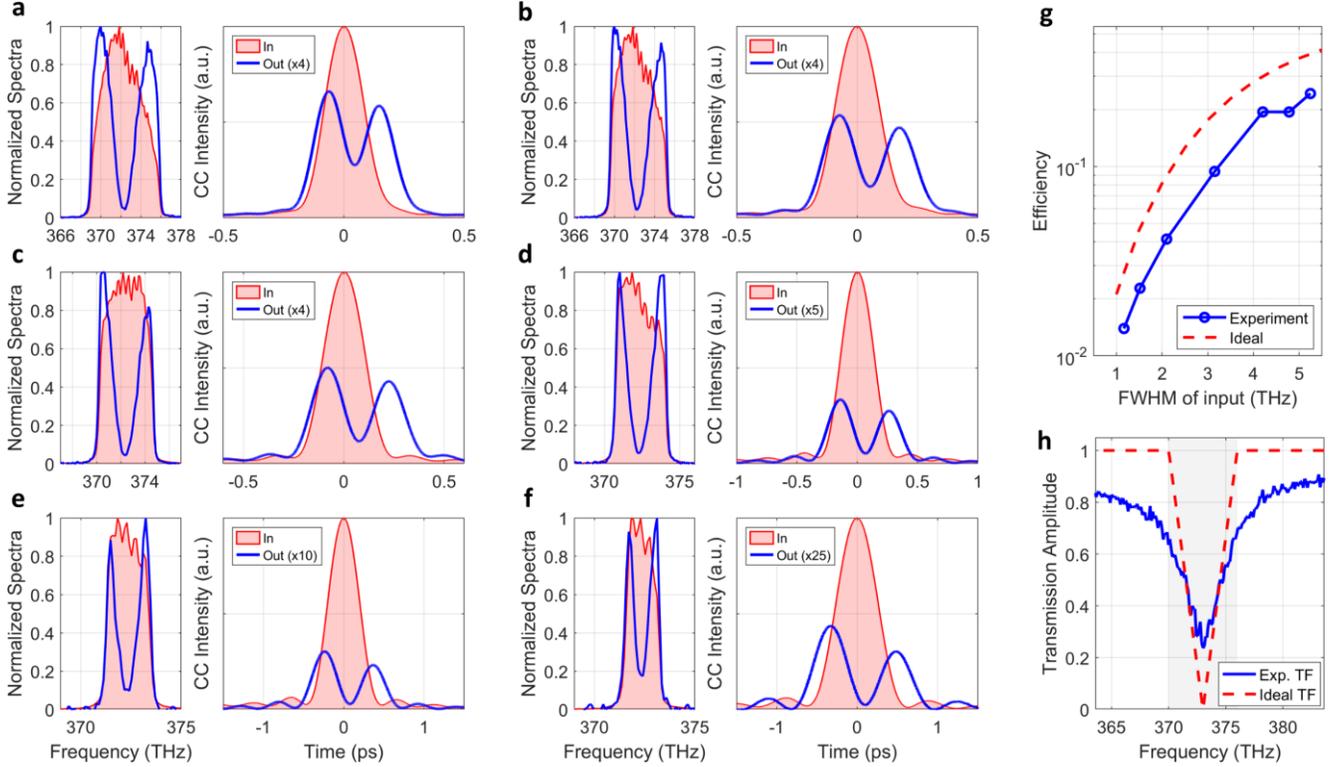

**Figure 4. Metasurface efficiency versus input bandwidth.** (a-f) Experimental measurements of the metasurface response for input pulses with different bandwidths. In each plot, the left panel shows the frequency spectrum of the input pulse (red-shaded area) and the filtered pulse (blue solid line). The right panel shows the corresponding CCTs. (g) Computational efficiency, defined as the ratio between the energies of the output and input pulses, versus the bandwidth of the input pulse, extracted for the experimental data (solid blue line) and calculated for an ideal differentiator with the same bandwidth as our metasurface. (h) Absolute value of the transfer function in the experiment (solid blue line), compared to the transfer function of an ideal differentiator with a bandwidth of 6 THz (dashed red line). See text for additional details.

In fact, the efficiency of our metasurface is close to the efficiency of an ideal passive differentiator device with the same bandwidth. To show this, we considered an ideal differentiator described by a transfer function $t_{ideal}(f) = 2(f - f_0)/\Delta f$ when $f \in [f_0 - \Delta f/2, f_0 + \Delta f/2]$, and $|t_{ideal}(f)| = 1$ otherwise. The central frequency $f_0 = 373$ THz and bandwidth $\Delta f = 6\ THz$ are chosen to match the ones of the measured device. The transfer function $|t_{ideal}(f)|$ is plotted in Fig.4h (dashed red line), together with the measured transfer function $|t_{meas}(f)|$ (solid blue lines). By using Gaussian pulses with different values of FWHMs as test input signals, we calculated the efficiency of such ideal differentiator as a function of the bandwidth of the input pulse. The results (red dashed line in Fig. 4f) confirm that the experimentally measured efficiency of our metasurface is less than a factor of 2 lower than the efficiency of an ideal differentiator with the same bandwidth.

Our nonlocal metasurface can perform high-efficiency first-order temporal differentiation of an optical signal within a subwavelength footprint ($H \approx \lambda_0/5$), in contrast with other metasurface-based implementations which rely on pulse shapers and bulky optics. The extreme compactness of our platform provides a natural opportunity to scale the optical computing tasks in a modular fashion, i.e., by cascading multiple elements. To demonstrate this feature, we performed an experiment where two metasurfaces, each separately performing first-order differentiation, are cascaded (Fig. 5a), and we verified that the operation imparted by the overall device is proportional to the second-order derivative of the input. The first metasurface (MS1) is the same as the one used for the experiments in Figs. 3 and 4. The second metasurface (MS2) had a nominally identical design as MS1, but fabricated separately. As input signal, we considered the same cosine-like signal used for the experiment in Fig. 3d (the corresponding CCT is shown in Fig. 5b as red-shaded area). We first tested each metasurface separately and confirmed that the output signals created by either MS1 or MS2 (solid and dashed blue lines in Fig. 5b) are almost identical, and they match the first-order derivative of the input signal (compare also with Fig. 3d). Then, we cascaded MS1 and MS2 (Fig. 5a, see Methods for more details on the setup) and we repeated the experiment with the same input signal. In the experiment (Fig. 5a), the metasurfaces were kept parallel to each other and separated by a distance of about 1 mm, but remarkably the output is not affected by the relative positioning of the two elements. Such a relatively large gap was only dictated by practical mechanical constraints of our setup, and it can be reduced to a few wavelengths without impacting the optical behavior. The measured output CCT (red solid line in Fig. 5c) displays the features expected from the second-order derivative of the cosine-like function used as input (red-shaded area in Fig. 5c). In particular, both input and output fields feature the same number of peaks, and with the same relative heights. The output is delayed by approximately $2\tau = 94$ fs with respect to the input, as expected from the two-metasurface arrangement. To confirm quantitatively that the output produced by the cascaded arrangement is the second-order derivative, we calculated the expected CCT for the exact second-order derivative of the input signal, following the same procedure as one used in Figs. 3(d-f). The result of this calculation, shown as dashed green line in Fig. 5c, matches almost perfectly the measured output CCT (red solid line).

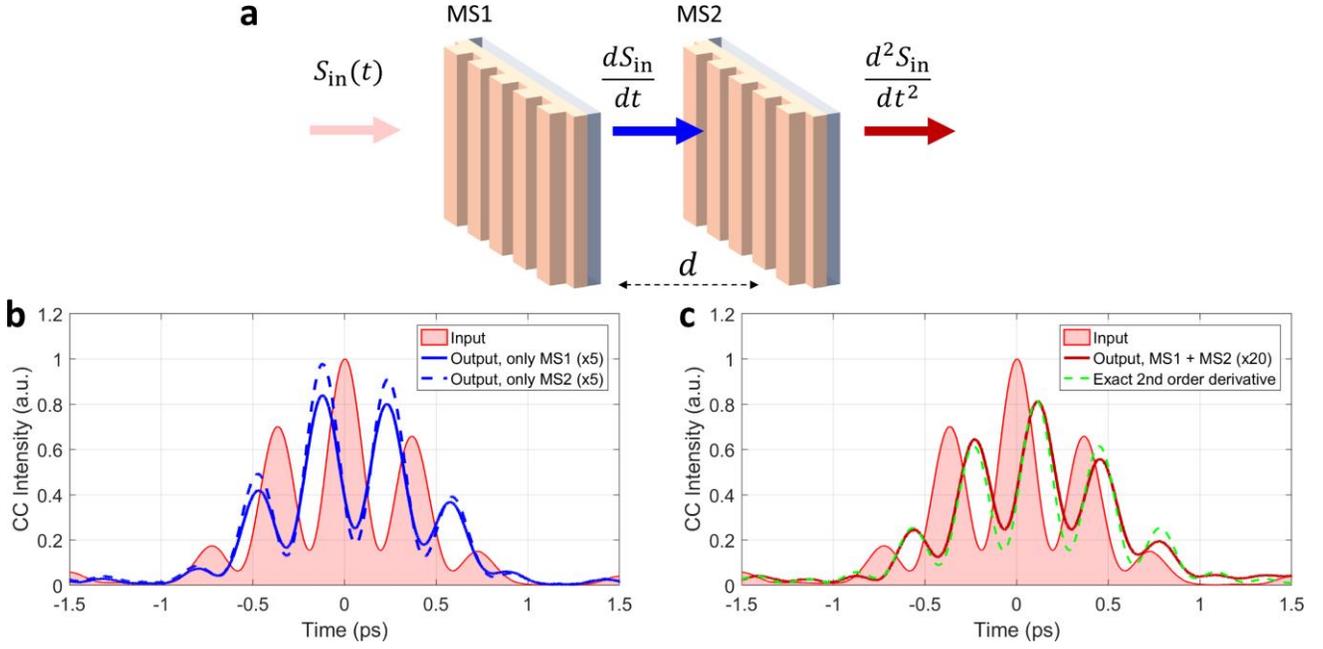

**Figure 5. Cascaded computational metasurfaces.** (a) Two similar metasurfaces, each of them performing first-order differentiation, are cascaded. The resulting device performs second-order differentiation. (b) Response of each metasurface. The input field (red shaded area) is the same as the one used in Fig. 3d. The solid and dashed blue lines show the CCT of the output fields produced by each metasurface separately. (c) Response of the cascaded system. The solid red line shows the output CCT when the input field impinging on the cascaded arrangement. The dashed green line shows the calculated CCT for the exact second-order derivative of the input field.

## Discussion and Conclusions

We have proposed and experimentally demonstrated a deeply subwavelength nonlocal metasurface that can perform analog first-order differentiation on the envelope of an optical signal. The functionality was achieved by engineering the temporal nonlocality of the metasurface, in order to tailor its frequency-dependent transfer function. Importantly, differently from approaches based on pulse shapers[9,10,41], in our experiment the computation is entirely performed by a single metasurface with thickness $H \approx \lambda_0/5$, and it does not require any additional optical element. We measured the metasurface response for different types of input pulses, confirming that the performed operation is very close to the exact first-order derivative. Our metasurface can be used to perform analog temporal edge detection, i.e., it can be used to identify the instant of times at which a signal is strongly varying. Moreover, thanks to the design simplicity and the absence of material loss, our metasurface features large integrated efficiency. Finally, we experimentally demonstrated that our device offers a natural platform to implement cascaded operations in a modular fashion, i.e., by physical cascading multiple metasurfaces. As a practical example, we showed that when two metasurfaces, each performing first-order derivative, are cascaded, the resulting device performs second-order differentiation.

Our results provide the first experimental demonstration that nonlocal metasurfaces can perform analog temporal operations on optical signals without the need of bulky optical setups, resulting in an extremely compact free-space computational block. The specific functionality demonstrated in this work — temporal edge detection — can be readily extended to metasurfaces with spatio-temporal nonlocalities to perform event-based detection[34] and analog neuromorphic computing. Moreover, the simplicity of the design introduced here makes it compatible with different tuning approaches[43,44], paving the way for reconfigurable computing with metasurfaces. Thanks to their reduced footprint, their bias-free operation and their ease to implementing scalable and cascaded operations, we envision that these devices will be adopted as building blocks within more complex optical computing platforms, especially in scenarios where low energy consumption and very low latency times are critical, such as neural networks and neuromorphic computing.

## Methods

**Numerical Simulations**

The metasurface geometry was optimized by performing numerical electromagnetic simulations with Comsol Multiphysics 6.1. The calculations in Fig. 1d were performed with a Matlab script, by calculating the Fourier transform of a given input signal, multiplying it by the metasurface transfer function calculated with Comsol (Fig. 1c), and then applying the inverse Fourier transform.

**Fabrication**

The metasurfaces were fabricated with a standard top-down lithographic process, starting with commercially available TiO2-coated glass substrates (MTI Corp). The TiO2 layer was first etched to the desired thickness $H = h + t$ (see Fig. 1c). A 20-nm-thick layer of chromium was deposited via e-beam evaporation, and a 300-nm-thick layer of e-beam resist (ZEP 520-A) was spin-coated on top of the samples. The metasurface pattern was written with an electron beam tool (Elionix 100 keV). After development, the pattern was transferred to the chromium layer via a Cl2−O2 dry etching process performed in an ICP machine (Oxford PlasmaPro System 100). After removing the ZEP mask, the pattern was transferred to the TiO2 layer with a CF4−Ar−O2 dry etching process performed in the same ICP machine. In this step, the TiO2 was etched to a depth equal to $h$ (see Fig. 1b). The residual chromium mask was removed via wet etching.

**Optical Characterization**

The normal-incidence transmission spectrum in Fig. 1f was measured with a custom-built setup. A collimated broadband lamp was weakly focused on the metasurface with a long-focal-length lens (f = 20 cm), and the beam transmitted through the sample was collected by an identical lens (L2) and sent onto the input slit of a visible spectrometer (Ocean Optics, HR 4000).

All time-domain measurements in Figs. 3-5 were performed with the setup shown in Fig. 2 and described in the text. To create pulses with different temporal profiles, we used spatial masks with different shapes in the pulse shaper. The masks required to create the beating pulses (Figs. 3(d-f) and Fig. 5) were obtained by drilling pairs of holes with diameter of 0.77 mm onto a thick metallic slab. To create the pulses with variable bandwidth (Fig. 4), we utilized a continuously variable mechanical slit (Thorlabs, VA100). To create rectangular pulses, we used a spatially varying mask that imparts a transmission profile proportional to $\sin(kx)/x$, where $x$ is a spatial coordinate along the mask and $k$ is a coefficient that determines the pulse duration. The mask was realized by cascading two masks, following the procedure described in ref. 45. The first mask, which imparts the transmission amplitude profile $\propto |\sin(kx)/x|$, consisted of metallic gratings with variable duty cycles. The second mask, which imparts the required binary phase profile $\propto \angle(\sin(kx)/x)$, was obtained by etching rectangular apertures onto a substrate of silicon oxide.

**Data Analysis**

In Figs. 3(d-f) and Fig. 5c we showed the numerically calculated output cross-correlation traces (CCTs) for the case in which the input fields are given by the beating of two beams with different frequencies. The CCTs were calculated by assuming that each metasurface performs the exact first-order derivative of the input signal, according to the following procedure. We assumed that the electric field pulse created by the pulse shaper is given by the sum of two gaussian pulses,

$$E_{in}(t) = \left( A_1 e^{-2\ln(2)\left(\frac{t}{\tau_1}\right)^2} e^{i2\pi\Delta \cdot t} + e^{i\phi} A_2 e^{-2\ln(2)\left(\frac{t}{\tau_2}\right)^2} e^{-i2\pi\Delta \cdot t} \right) e^{i2\pi f_0 t}.$$

The central frequency $f_0$ and the detuning between the beams $\Delta$ can be retrieved from independent spectroscopic measurements. The parameters $\tau_1$ and $\tau_2$ describe the bandwidth of each beam, and they are mainly determined by the width of the pinholes in the mask. The parameters $A_1$ and $A_2$ describe the relative amplitudes of each beam, and only their ratio $A_1/A_2$ is of practical importance here. The narrow reference pulse (used in the experiment to produce the CCTs) is described by

$$E_{ref}(t) = e^{-2\ln(2)\left(\frac{t}{\tau_{ref}}\right)^2} e^{i2\pi f_0 t},$$

where we assume $\tau_{ref} = 80\,fs$ based on the independent characterization of the reference pulse. We used this ansatz to calculate the expected CCT of the input field, $V_{in}^{calc}(\tau) = \int dt\, I_{in}(t) I_{ref}(\tau - t)$, where $I_{in}(t) = |E_{in}(t)|^2$ and $I_{ref}(t) = |E_{ref}(t)|^2$. By comparing the calculated $V_{in}^{calc}(\tau)$ with the measured CCT of the input field, we were able to estimate with high accuracy all remaining parameters in the expression for $E_{in}(t)$. In particular, for all cases considered here we found that the experimental data are well fitted by assuming that the two beams have the same amplitude ($A_1 = A_2$), the same bandwidth ($\tau_1 = \tau_2 \approx 0.98$ ps) and the same phase ($\phi = 0$). After having fully determined $E_{in}(t)$, we calculated numerically the first-order derivative of its envelope, i.e. $g(t) = \frac{d}{dt}[E_{in}(t)e^{-i2\pi f_0 t}]$, and we used it to construct the expected output electric field $E_{out}(t) = g(t)e^{i2\pi f_0 t}$. Finally, we calculated the expected CCT of the output field, $V_{out}^{calc}(\tau) = \int dt\, I_{out}(t) I_{ref}(\tau - t)$, where $I_{out}(t) = |E_{out}(t)|^2$. The green dashed lines in Figs. 3(d-f) correspond to the curve $V_{out}^{calc}(\tau)$ calculated with this procedure and rescaled vertically by an overall factor. For the case of the second-order derivative (green dashed lines in Fig. 5c) we followed the same procedure outlined above, but now $g(t) = \frac{d^2}{dt^2}[E_{in}(t)e^{-i2\pi f_0 t}]$.

## Funding

## Acknowledgment

All authors conceived the idea and the corresponding experiment. S.E. and M.C. designed the metasurface and performed the numerical calculations. M.C. fabricated the samples. D.K. built the setup. M.C. and D.K. performed all measurements with assistance from S.E.. A.A. supervised the project. All authors analyzed the data and contributed to writing the manuscript. Device fabrication was performed at the Nanofabrication Facility at the Advanced Science Research Center at The Graduate Center of the City University of New York.

## Disclosures

The authors declare no conflicts of interest.

## Data availability

Data underlying the results presented in this paper may be obtained from the authors upon reasonable request.